\documentclass[12pt,preprint]{emulateapj}
\usepackage{graphicx}

\shorttitle{Phase Variations of Eccentric Planets}
\shortauthors{Stephen R. Kane \& Dawn M. Gelino}
\slugcomment{Submitted for publication in the Astrophysical Journal}

\begin{document}

\title{Photometric Phase Variations of Long-Period Eccentric Planets}
\author{Stephen R. Kane, Dawn M. Gelino}
\affil{NASA Exoplanet Science Institute, Caltech, MS 100-22, 770
  South Wilson Avenue, Pasadena, CA 91125, USA}
\email{skane@ipac.caltech.edu}


\begin{abstract}

The field of exoplanetary science has diversified rapidly over recent
years as the field has progressed from exoplanet detection to
exoplanet characterization. For those planets known to transit, the
primary transit and secondary eclipse observations have a high yield
of information regarding planetary structure and atmospheres. The
current restriction of these information sources to short-period
planets may be abated in part through refinement of orbital
parameters. This allows precision targeting of transit windows and
phase variations which constrain the dynamics of the orbit and the
geometric albedo of the atmosphere. Here we describe the expected
phase function variations at optical wavelengths for long-period
planets, particularly those in the high-eccentricity regime and
multiple systems in resonant and non-coplanar orbits. We apply this to
the known exoplanets and discuss detection prospects and how
observations of these signatures may be optimized by refining the
orbital parameters.

\end{abstract}

\keywords{planetary systems -- techniques: photometric -- techniques:
  radial velocities}


\section{Introduction}
\label{introduction}

The currently known diversity of exoplanets is greatly attributable to
the revolution of the transit detection method over the past 10 years.
The measurement of radius and hence density were the first steps from
the results of this technique, but soon to follow were atmospheric
studies from both primary transit and secondary eclipse. However, the
unknown inclination of the planetary orbits makes this technique only
applicable to a relatively small fraction of the known exoplanets. For
the non-transiting planets, reflected light and phase variations
present an additional avenue through which to investigate planetary
atmospheres \citep{cha99,lei03}. The net result of this new
information has lead to an unprecedented ability to characterize
exoplanets.

Phase functions in the Infra-Red (IR) primarily measure the thermal
properties of the planet whereas optical measurements probe the
planetary albedo. The relation between giant planet atmospheres and
phase curves have been described in detail by
\citet{sud05}. \citet{iro10} further investigate the time-variation of
the atmospheres in the IR for eccentric planets using radiative
transfer models. The phase variation of our own solar system has been
investigated by \citet{dyu05}, and it has been shown that phase
functions can be used to produce longitudinal thermal maps of
exoplanets \citep{cow08}. In addition, phase curves of terrestrial
planets have been considered by \citet{mal09}.

Phase variations of exoplanets in the IR and optical regimes have had
success due to increased access to improved instrumentation and
space-based observatories. This has primarily been investigated for
transiting planets since the edge-on orbital plane produces the
highest phase amplitude. Examples of observed phase variations in the
IR (using Spitzer) include HD~189733b \citep{knu09a} and HD~149026b
\citep{knu09b}. Examples in the optical include Kepler observations of
HAT-P-7b \citep{wel10} and phase variations detected in the light
curve of CoRoT-1b \citep{sne09}.

\citet{kan08} and \citet{kan09a} showed that planets in eccentric
orbits have inflated transit probabilities, as demonstrated by
HD~17156b \citep{bar07} and HD~80606b \citep{lau09}. These types of
planets will produce relatively high phase amplitudes during a brief
period (periaston passage) of the orbit. However, phase variations of
non-transiting planets have been restricted to hot Jupiters, including
$\upsilon$~And~b \citep{har06} and HD~179949b \citep{cow07}. There
have been searches for phase variations of, for example, HD~75289Ab
\citep{rod08} and $\tau$~Boo~b \citep{cha99,rod10} but no signatures
were detected in either case. Exploring the atmospheric properties of
longer-period planets requires taking advantage of highly-eccentric
non-transiting systems. Thus we also constrain the inclination and
hence the mass of such planets.

Here we investigate the expected photometric phase amplitude of
long-period eccentric planets and show how planetary orbits in
resonance can result in ambiguous phase variation detections. We
further apply this analysis by calculating maximum flux ratios for the
known exoplanets and considering several interesting case studies. We
also determine the effective orbital phase regime over which
detectability is maximised and show how refinement of orbital
parameters can allow efficient targeted observations at these
times. This study is intended from an observers point of view in so
far as these observable variations can be reconnected back to the
theoretical models of exoplanetary atmospheres.


\section{Exoplanet Phase Variations}

\begin{figure}
  \includegraphics[width=8.2cm]{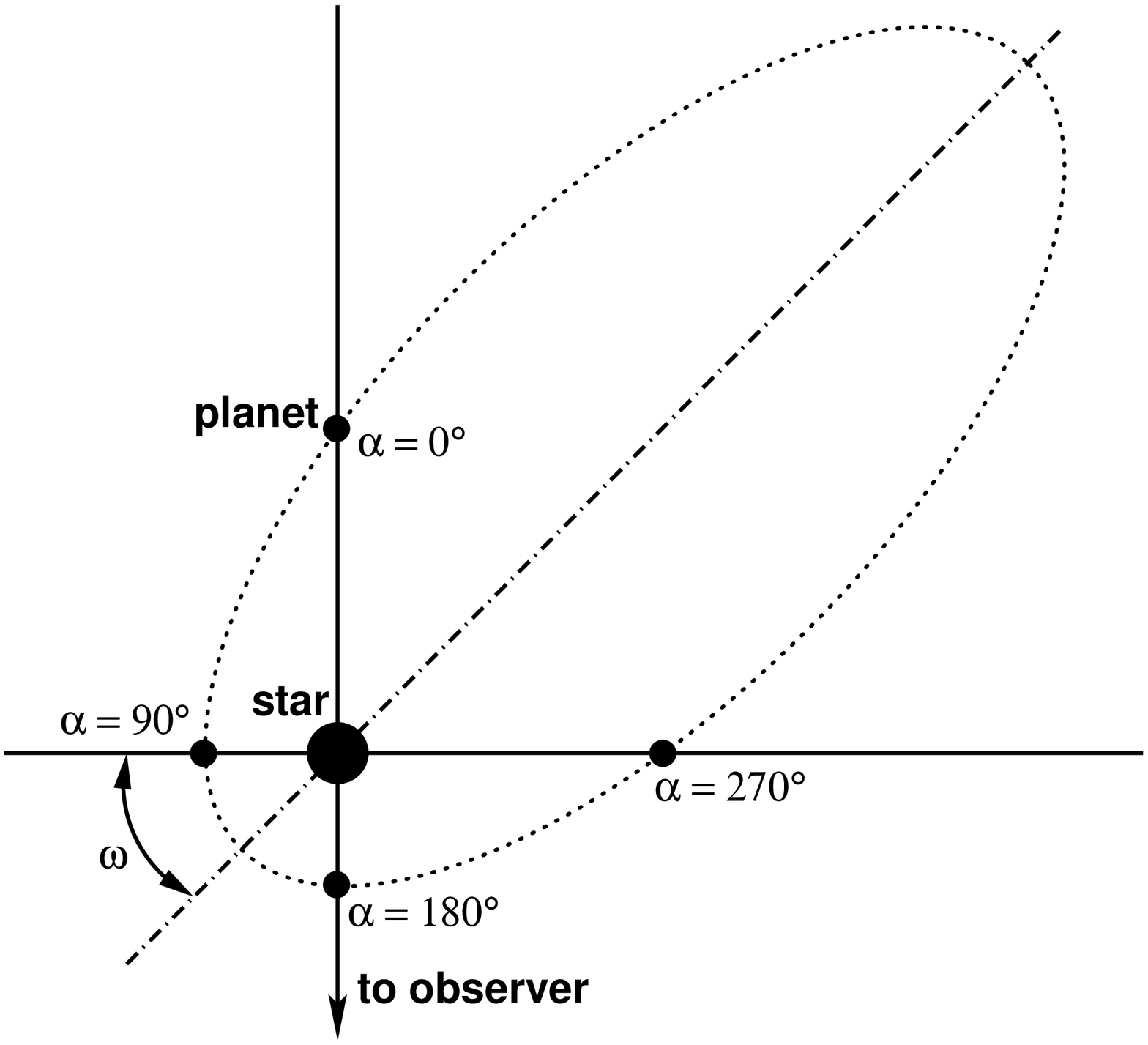}
  \caption{Orbit of eccentric planet, showing orbit phase angles
    corresponding to full ($\alpha = 0\degr$), first quarter ($\alpha =
    90\degr$), new ($\alpha = 180\degr$), and third quarter ($\alpha =
    270\degr$) phases.}
  \label{pha_orbit}
\end{figure}

In this section, we establish the theoretical framework which will be
applied in the remainder of the paper, similar to the formalism used
by \citet{col02} and more recently by \citet{rod10}. Figure
\ref{pha_orbit} shows a top-down view of an elliptical planetary
orbit. The phase angle $\alpha$ is described by
\begin{equation}
  \cos \alpha = \sin (\omega + f)
  \label{phaseangle}
\end{equation}
where $\omega$ is the argument of periastron and $f$ is the true
anomaly. The phase angle is defined to be $\alpha = 0\degr$ when the
planet is at superior conjunction (``full'' phase). In terms of
orbital parameters, this location in the orbit occurs when $\omega + f
= 270\degr$.

The flux at wavelength $\lambda$ incident upon the planet is described
by
\begin{equation}
  F_i(\lambda) = \frac{L_\star(\lambda)}{4 \pi r^2}
\end{equation}
where $L_\star$ is the luminosity of the star and $r$ is the
star--planet separation. This separation is given by
\begin{equation}
  r = \frac{a (1 - e^2)}{1 + e \cos f}
  \label{separation}
\end{equation}
where $a$ is the semi-major axis and $e$ is the orbital eccentricity.
The geometric albedo of a planet is defined at $\alpha = 0\degr$ as
follows
\begin{equation}
  A_g(\lambda) = \frac{F_r(0,\lambda)}{F_i(\lambda)}
\end{equation}
where $F_r$ is the reflected light from the planet.
The planetary flux received at Earth is then
\begin{equation}
  f_p(\alpha,\lambda) = A_g(\lambda) g(\alpha,\lambda) F_i(\lambda)
  \frac{R_p^2}{d^2}
\end{equation}
where $R_p$ is the planetary radius, $d$ is the distance to the star,
and $g(\alpha,\lambda)$ is the phase function. Since the stellar flux
received at Earth is
\begin{equation}
  f_\star(\lambda) = \frac{L_\star(\lambda)}{4 \pi d^2}
\end{equation}
then the flux ratio of the planet to the host star is defined as
\begin{equation}
  \epsilon(\alpha,\lambda) \equiv
  \frac{f_p(\alpha,\lambda)}{f_\star(\lambda)}
  = A_g(\lambda) g(\alpha,\lambda) \frac{R_p^2}{r^2}
  \label{fluxratio}
\end{equation}
and thus contains three major components; the geometric albedo, the
phase function, and the inverse-square relation to the star--planet
separation. Note that for a circular orbit, only the phase function is
time dependent.


\subsection{Wavelength Dependence}
\label{wavelength}

As noted in the previous section, the observed flux ratio from an
exoplanet is wavelength dependent. In particular, the atmospheric
composition drives the scattering properties and thus the forms of the
geometric albedo and phase function. This dependence has been
considered in great detail by \citet{sud05} in which they construct
empirical models for exoplanet atmospheres and intergrate over the
surface with assumed opacities depending upon atmospheric
composition. This thorough analysis is not reproduced here, but we do
use the results of their analysis for a restricted wavelength,
particularly with regards to the albedo function discuessed in the
following sections. Here we confine our study to optical wavelengths
centered on 550~nm. This broadly encompasses the results obtained by
such studies at \citet{col02}, \citet{lei03}, and \citet{rod10}. This
also places the study near the peak response of the Kepler CCD, the
relevance of which will be discussed in later sections.


\subsection{Geometric Albedo}

It has been shown through atmospheric models that there is a
dependence of the geometric albedo of giant planets on the semi-major
axis of the orbit \citep{sud00,sud05,cah10}. We direct the reader to Figure
9 of \citet{sud05} which details the wavelength and star--planet
separation dependence of the geometric albedo. Jupiter is known to
have a visual geometric albedo of $\sim 0.52$. However, the strong
irradition of the atmospheres of giant planets in short-period orbits
results in the removal of reflective condensates from the upper
atmospheres and thus a significant lowering of the geometric
albedo. Observations of HD~209458b using the Microvariability and
Oscillations of STars (MOST) satellite by \citet{row08} failed to
detect phase variations and thus they were able to place an upper
limit of $A_g < 0.08$, subsequently investigated using model
atmospheres by \citet{bur08}. More recent observations of HAT-P-7b
using Kepler by \citet{wel10} revealed phase variations in the light
curve from which they were able to deduce a geomtric albedo of 0.18.

\begin{figure}
  \includegraphics[angle=270,width=8.2cm]{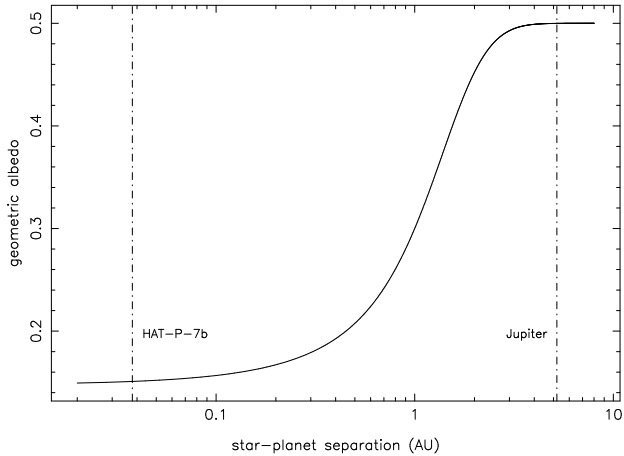}
  \caption{Approximation of the geometric albedo distribution for
    giant planets, where $\lambda \sim 550$~nm.}
  \label{albedofig}
\end{figure}

Through the examples mentioned above, and the consideration of the
theoretical models of \citet{sud05}, we have constructed a model which
approximates the albedo of giant planets as a function of star--planet
separation (see Equation \ref{separation}). This is a hyperbolic
tangential function of the form
\begin{equation}
  A_g = \frac{(e^{r-1}-e^{-(r-1)})}{5(e^{r-1}+e^{-(r-1)})} +
  \frac{3}{10}
  \label{albedo}
\end{equation}
Equation \ref{albedo} is plotted in Figure \ref{albedofig}, showing
the semi-major axes of HAT-P-7b and Jupiter for reference. This
function well represents the rapid rise in optical albedo between 0.2
and 1 AU described by \citet{sud05}, as well as the continued rise
beyond 2 AU whereby water clouds begin to be present. Though broadly
robust to encompass both the theoretical calculations and the limited
number of measured examples mentioned above, this empirical function
does not account for planetary gravity whose variation may affect the
atmospheric properties and thus the albedo properties.

\subsection{Phase Function}

A planetary phase function can be considered to be defined by the
continuous presence of an imaginary line conecting the center of the
star and planet which is normal to the day--night terminator of the
planet. The phase function of a Lambert sphere assumes the atmosphere
isotropically scatters over $2 \pi$ steradians and is described by
\begin{equation}
  g(\alpha,\lambda) = \frac{\sin \alpha + (\pi - \alpha) \cos
    \alpha}{\pi}
\end{equation}
and is thus normalized to lie between 0 and 1.  For a circular orbit,
the phase function applied this to the flux ratio relation (Equation
\ref{fluxratio}) results in both a phase function and flux ratio which
are maximum at a phase angle of zero. Generalizing the phase angle
(see Equation \ref{phaseangle}) and thus the phase function for an
eccentric orbit requires first solving Kepler's equation
\begin{equation}
  M = E - e \sin E
\end{equation}
where $M$ is the mean anomaly and $E$ is the eccentric anomaly. The
true anomaly is then related to the eccentric anomaly by
\begin{equation}
  \cos f = \frac{\cos E - e}{1 - e \cos E}
\end{equation}
where the true anomaly establishes the time-dependent variation of the
phase function.

\begin{figure*}
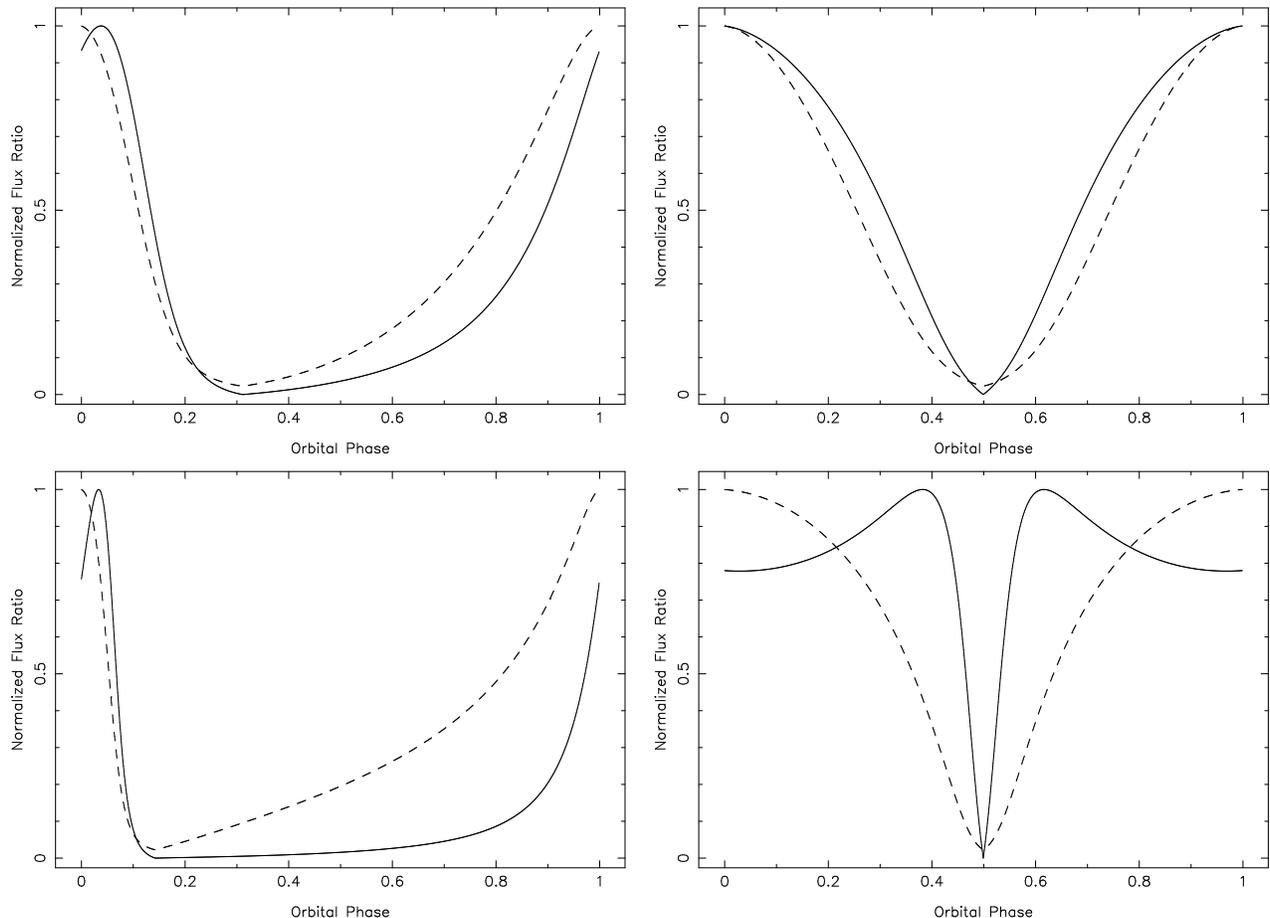

  \begin{center}
    \begin{tabular}{cc}
      \includegraphics[angle=270,width=8.2cm]{f03a.ps} &
      \includegraphics[angle=270,width=8.2cm]{f03b.ps} \\
      \includegraphics[angle=270,width=8.2cm]{f03c.ps} &
      \includegraphics[angle=270,width=8.2cm]{f03d.ps} \\
    \end{tabular}
  \end{center}
  \caption{The phase functions (dashed line) and normalized flux
    ratios (solid line) for various eccentricities and periastron
    arguments; $e = 0.3$ and $\omega = 0\degr$ (top-left), $e = 0.3$
    and $\omega = 90\degr$ (top-right), $e = 0.6$ and $\omega =
    0\degr$ (bottom-left), $e = 0.6$ and $\omega = 90\degr$
    (bottom-right).}
  \label{pha_ecc}
\end{figure*}

For the analysis performed here, we adopt the approach of
\citet{col02} and \citet{rod10} which utilizes the empirically derived
phase function of \citet{hil92}. This is based upon observations of
Jupiter and Venus and incorporates substantially more back-scattering
due to cloud-covering. This approach contains a correction to the
planetary visual magnitude of the form
\begin{equation}
  \Delta m (\alpha) = 0.09 (\alpha/100\degr) + 2.39
  (\alpha/100\degr)^2 -0.65 (\alpha/100\degr)^3
  \label{magcorr}
\end{equation}
which leads to a phase function given by
\begin{equation}
  g(\alpha) = 10^{-0.4 \Delta m (\alpha)}
  \label{phase}
\end{equation}
where the wavelength dependence has been removed (see Section
\ref{wavelength}). This Hilton phase function is used throughout the
remainder of this paper.

Shown in Figure \ref{pha_ecc} are phase functions and
normalized flux ratios for various eccentricities and orbital
orientations. Note that the maximum flux ratio does not necessarily
occur at zero phase angle for a non-circular orbit. This is because
the orbital distance is changing and indeed we shall show in later
sections that the star--planet separation component of Equation
\ref{fluxratio} becomes dominant for highly eccentric orbits. This
time-lag between maximum flux ratio and maximum phase was also noted
by \citet{sud05}.


\subsection{Orbital Inclination}
\label{inclination}

For interacting systems, many of the system parameters depend on the
orbital inclination angle, $i$, of the system (for example, see
\citet{gel06}). For exoplanetary systems, given an assumed albedo, the
true amplitude of the phase variation can be used to estimate the
inclination angle, and therefore constrain the mass of the planet
derived from radial velocity data. To add the effect of inclination
angle to the phase function, the phase angle (Equation
\ref{phaseangle}) is modified as follows:
\begin{equation}
  \cos \alpha = \sin (\omega + f) \sin i
  \label{phaseinc}
\end{equation}
At 1st and 3rd quarter ($\alpha = 90\degr$ and $\alpha = 270\degr$),
the flux ratio is completely independent of inclination angle (see
Figure \ref{pha_orbit}). The effect of inclination on the shape of the
phase function is quite small and has negligible effect on the
location of the minimum and maximum values of the flux ratio, as shown
by Figure 20 of \citet{sud05}. One complicating factor in this simple
inclination consideration is that any additional light sources
(i.e. planets) in a given system will dilute the signature from the
dominant light-reflecting planet. The dilution of the signature will
remain constant if the additional planets are in face-on ($i =
0\degr$) orbits. This will be discussed further in Section
\ref{coplanarity}.


\section{Multi-Planet Systems}

Detection of multi-planet systems is becoming more frequent as we are
increasingly able to probe into smaller mass and longer period regimes
of parameter space. If indeed single planet systems are rare, then it
is highly likely that the phase curve from a particular planet will be
``contaminated'' by the reflected light of other planets in the
system. For planets which are similar in size, the effect of an outer
planet to the combined phase curve will be small since (from Equation
\ref{fluxratio}) $\epsilon(\alpha,\lambda) \propto r^{-2}$. Here we
discuss the specific cases of orbital coplanarity and resonant orbits.


\subsection{Coplanarity}
\label{coplanarity}

Additional planets in a system serve to dilute the signature from the
dominant light-reflecting planet. Depending upon planet formation
scenarios, it cannot be taken for granted that planets within a system
will lie in coplanar orbits. As shown in Section \ref{inclination},
the phase function may be reduced in amplitude significantly for
orbits inclined relative to the line-of-sight, to the extreme of
eliminating a time-variable photometric signature of the orbit if it
is face-on ($i = 0\degr$).

Consider the case of the planetary system orbiting the star
$\upsilon$~And, which was first discovered by \citet{but99}. A search
for reflected light from the inner-most planet was carried out by
\citet{col02}, but only a marginal detection was produced leaving an
ambiguity of the result concerning the degeneracy between the
planetary radii and assumed albedos. The $\upsilon$~And system is one
of the few systems which has been monitored astrometrically as well as
spectroscopically in order to provide constraints on the orbital
inclinations of the planets \citep{mca10}. The inclinations of the
outer two planets (c and d), with semi-major axes of 0.83~AU and
2.53~AU, were measured to be $\sim 8\degr$ and $24\degr$
respectively. The factor of 3 increase in orbital distance of the
outer planet results in a factor of 9 less contribution to the total
planetary reflected light from the system. However, the relative
inclinations of the planets causes the phase function of the outer
planet to be almost 3 times stronger than that of the inner planet.


\subsection{Resonance}

A number of systems have now been found to contain planets in
eccentric orbits with some kind of resonant behaviour. An interesting
example is the HD~82943 planetary system (see Section \ref{hd82943}),
the 2:1 resonance of which has been studied in detail by
\citet{lee06}. The periodic simulataneous periastron passage of two
planets produces a distinct signature from a phase amplitude
perspective, although those moments will not necessarily be those of
maximum flux ratio. In the context of radial velocity measurements,
\citet{giu09} describe how resonant orbits can affect the
detectability of exoplanets. Furthermore, it was shown by
\citet{ang10} that 2:1 resonant systems can be mis-interpreted as
single-planet eccentric orbits when performing a fit to the radial
velocity data. The same is true for phase curves of multi-planet
systems where resonant orbits can effectively hide the presence of the
outer planet in the resulting phase curve since the combined phase
variation will be periodic with time. In contrast, non-resonant
planets will in general produce a non-periodic combined signal that
will resolve as two separate phase functions with time.

\begin{figure*}
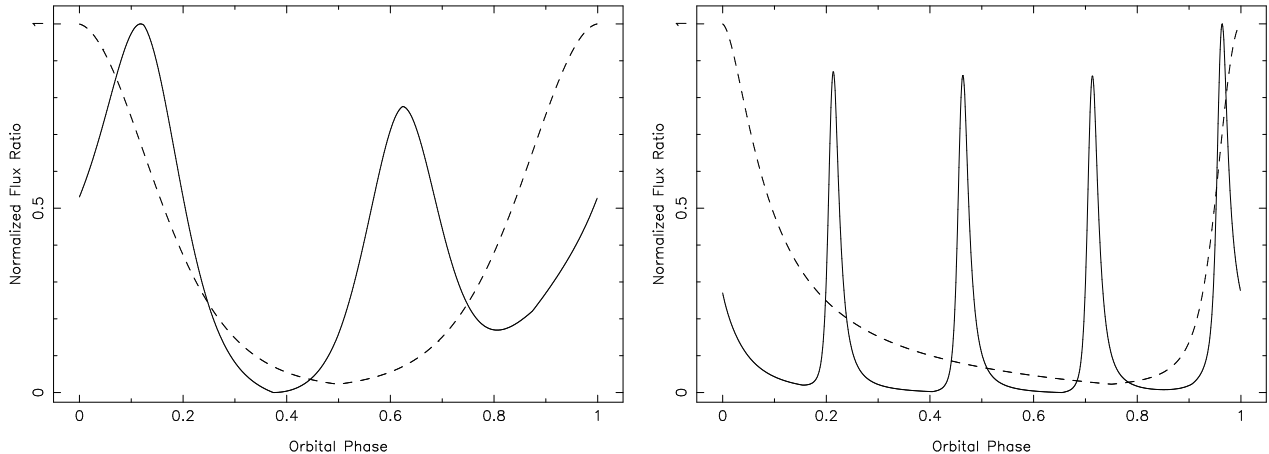

  \begin{center}
    \begin{tabular}{cc}
      \includegraphics[angle=270,width=8.2cm]{f04a.ps} &
      \includegraphics[angle=270,width=8.2cm]{f04b.ps}
    \end{tabular}
  \end{center}
  \caption{Normalized flux ratio (solid line) for a multi-planet
    system in which both planets are in 2:1 resonance with $e = 0.0$
    (left panel) and 4:1 resonance with $e = 0.5$ (right panel). The
    dashed line in each case represents the phase function of the
    outer planet.}
  \label{resonance}
\end{figure*}

Shown in Figure \ref{resonance} are two example systems, each with two
Jupiter radii planets. The orbits of the first system are in 2:1
resonance with $e = 0.0$ while the planets of the second system are in
4:1 resonance with $e = 0.5$. Given sufficient photometric precision
and observing cadence, it may be possible to distinguish the deviant
secondary peak of the first system and deduce the presence of the
outer planet. However, this will be a difficult endeavour since the
two peaks are relatively similar in amplitude. The second system
presents an even more difficult problem, with the combination of high
eccentricity and larger relative semi-major axis of the outer planet
leading to a limited observation window, higher required cadence, and
modest phase signature from the outer planet (seen close to an orbital
phase of 0.95 in Figure \ref{resonance}). If one is unable to monitor
this highest peak and also discern the difference in amplitude with
the other three peaks then the presence of the outer planet will
remain hidden to the observer. Thus the derived system architecture
based purely upon the phase variations will be incorrect. Resonant
systems such as these currently comprise a small fraction of the total
number of exoplanet systems. The relevance of this issue will increase
as radial velocity surveys sample to longer periods and as Kepler
discovers multi-planet systems, of which candidates have already been
announced \citep{ste10}.


\section{Application to Known Exoplanets}

Here we apply the results of the previous sections to the known
exoplanets. The orbital parameters of 370 planets were extracted using
the Exoplanets Data Explorer\footnote{\tt http://exoplanets.org/}. The
data are current as of 22nd May 2010. Since the flux ratio is $\propto
R_p^2$, the unknown planetary radii for the non-transiting planets
injects a degree of uncertainty into these calculations. The models of
\citet{bod03} and \citet{for07} show that there is a clear planetary
radius dependence upon stellar age as well as incident flux and
planetary composition. However, \citet{for07} also showed that, for a
given planetary composition, planetary radii should not vary
substantially between orbital radii of 0.1--2.0~AU. Since most of the
planets we are considering here lie beyond 0.1~AU from their parent
stars (by virtue of their eccentricity) and the mass distribution
peaks at one Jupiter mass in this region, we fix the radius for each
of the planets in this sample at one Jupiter radius, with the caveats
mentioned above in mind.

As shown earlier, the orbital phase at which the maximum flux ratio,
$\epsilon_{\mathrm{max}}$, occurs for an eccentric orbit depends upon
the orbital orientation. Since Kepler's equation is a transcendental
function, the integral of Equation \ref{fluxratio} must be solved
numerically in order to determine the maximum flux ratio for each
planet. These calculations assume an orbital inclination of $i =
90\degr$ and thus the maximum flux ratios are upper limits in most
cases, even though the radial velocity technique is biased towards
detection of edge-on orbits since these produce larger radial velocity
semi-amplitude signatures. We have also calculated the minimum time
difference in units of orbital phase between where maximum flux occurs
and where the flux drops to less than 5\% of the difference between
maximum and minimum flux. This quantity is designated $\Delta t$ and
represents the minimum time over which observations of maximum
effectiveness can be made.

\begin{figure*}
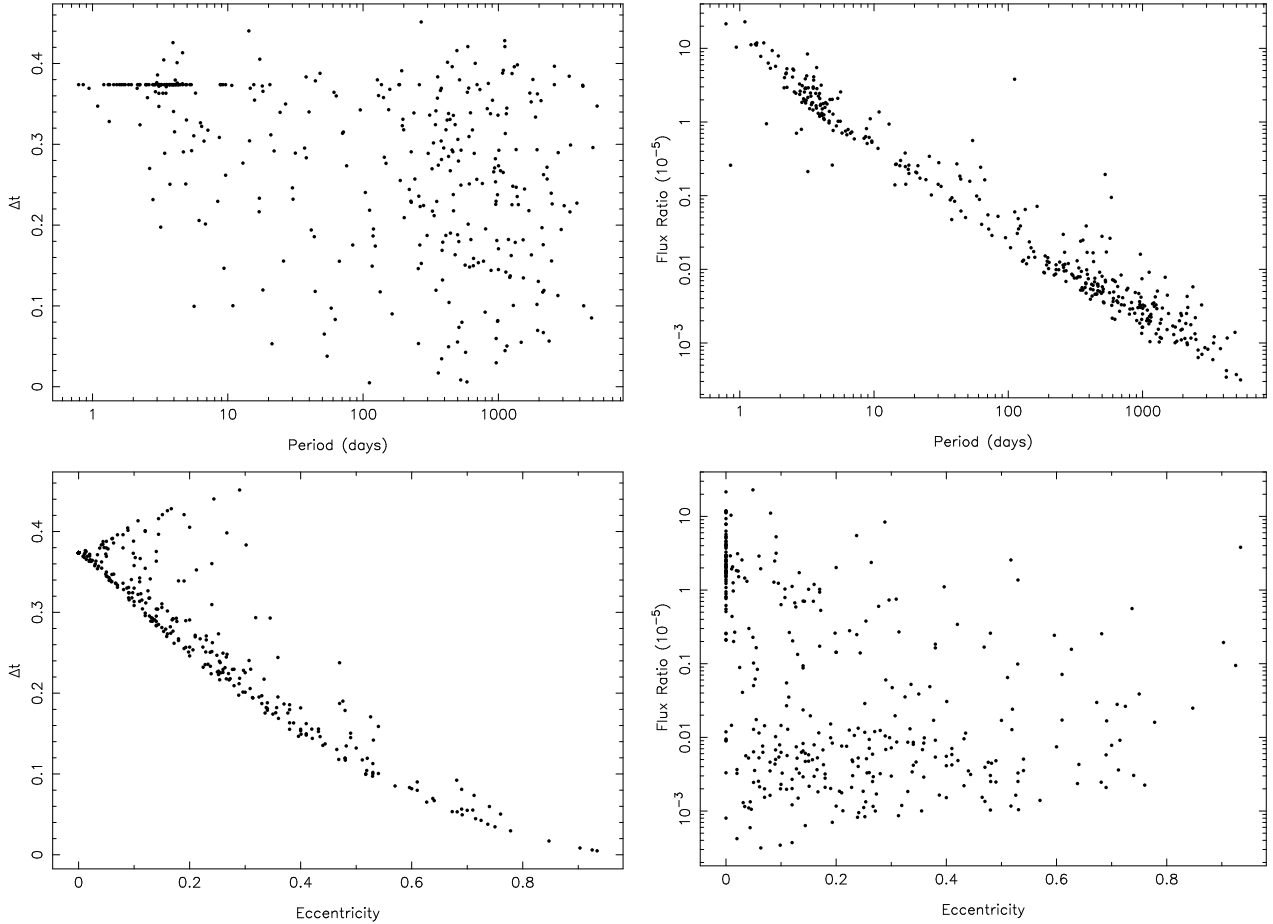

  \begin{center}
    \begin{tabular}{cc}
      \includegraphics[angle=270,width=8.2cm]{f05a.ps} &
      \includegraphics[angle=270,width=8.2cm]{f05b.ps} \\
      \includegraphics[angle=270,width=8.2cm]{f05c.ps} &
      \includegraphics[angle=270,width=8.2cm]{f05d.ps} \\
    \end{tabular}
  \end{center}
  \caption{Calculated values of maximum phase amplitude
    $\epsilon_{\mathrm{max}}$ and approximate time (in phase) between
    minimum and maximum amplitude $\Delta t$ for the 370 known
    exoplanets included in this sample. The top two panels show these
    calculated quantities plotted as a function of orbital period, and
    the bottom two show their variation as a function of orbital
    eccentricity.}
  \label{knownexo}
\end{figure*}

These calculated values are plotted in Figure \ref{knownexo}, both as
a function of period and eccentricity. The value of $\Delta t$ is
$\sim 0.37$ for all planets in circular orbits since this is where the
phase difference for a simple cosine variation crosses the $< 5$\%
threshold described above. For eccentric orbits $\Delta t$
can be larger than expected, particularly where $\omega \sim
270\degr$. Even so, the distribution shown in the top-left panel of
Figure \ref{knownexo} mirrors the distribution of orbital
eccentricities. The bottom-left panel shows that there is a minimum
value of $\Delta t$ that may occur for a given eccentricity, but once
again we see that this can float upwards depending upon the value of
$\omega$. The evident linear relation in log-space of the flux ratio
on period shown in the top-right panel demonstrates that the flux
ratio is indeed dominated by the star-planet separation as one would
expect. However, note the significant outliers beyond a period, $P$,
of 200 days which are caused by the highly eccentric planets which
pass through periastron close to a phase angle of $0\degr$. Several of
these systems are discussed in detail in Section \ref{casestudies}.

\begin{deluxetable}{lccccc}
  \tablecolumns{6}
  \tablewidth{0pc}
  \tablecaption{\label{phasetable} $\epsilon_{\mathrm{max}}$ and
    $\Delta t$ for eccentric exoplanets.}
  \tablehead{
    \colhead{Planet} & \colhead{$P$ (d)} & \colhead{$e$} &
    \colhead{$\omega$ ($\degr$)} & \colhead{$\Delta t$} &
    \colhead{$\epsilon_{\mathrm{max}} (10^{-5})$}
  }
  \startdata
HD 80606 b      &    111.43 &  0.93 &  300.60 & 0.005 &  3.8029 \\
HD 20782 b      &    585.86 &  0.93 &  147.00 & 0.006 &  0.0946 \\
HD 4113 b       &    526.62 &  0.90 &  317.70 & 0.008 &  0.1939 \\
HD 156846 b     &    359.51 &  0.85 &   52.23 & 0.017 &  0.0249 \\
HD 45350 b      &    963.60 &  0.78 &  343.40 & 0.030 &  0.0160 \\
HD 30562 b      &   1157.00 &  0.76 &   81.00 & 0.050 &  0.0022 \\
HD 20868 b      &    380.85 &  0.75 &  356.20 & 0.035 &  0.0388 \\
HD 41004 A b    &    963.00 &  0.74 &   97.00 & 0.060 &  0.0030 \\
HD 37605 b      &     54.23 &  0.74 &  211.60 & 0.038 &  0.5608 \\
HD 222582 b     &    572.38 &  0.73 &  319.01 & 0.043 &  0.0263 \\
HD 2039 b       &   1120.00 &  0.71 &  344.10 & 0.045 &  0.0091 \\
iota Dra b      &    511.10 &  0.71 &   91.58 & 0.073 &  0.0036 \\
HD 96167 b      &    498.90 &  0.71 &  285.00 & 0.055 &  0.0281 \\
HD 86264 b      &   1475.00 &  0.70 &  306.00 & 0.055 &  0.0078 \\
HAT-P-13 c      &    428.50 &  0.69 &  176.70 & 0.050 &  0.0167 \\
HD 159868 b     &    986.00 &  0.69 &   97.00 & 0.081 &  0.0021 \\
HD 43848 b      &   2371.00 &  0.69 &  229.00 & 0.056 &  0.0058 \\
HD 17156 b      &     21.22 &  0.68 &  121.90 & 0.053 &  0.2549 \\
16 Cyg B b      &    798.50 &  0.68 &   85.80 & 0.092 &  0.0025 \\
HD 89744 b      &    256.78 &  0.67 &  195.10 & 0.053 &  0.0298 \\
HD 39091 b      &   2151.00 &  0.64 &  330.24 & 0.067 &  0.0043 \\
HD 131664 b     &   1951.00 &  0.64 &  149.70 & 0.070 &  0.0024 \\
HD 74156 b      &     51.65 &  0.63 &  176.50 & 0.065 &  0.1567 \\
HD 171028 b     &    538.00 &  0.61 &  305.00 & 0.080 &  0.0172 \\
HD 154672 b     &    163.94 &  0.61 &  265.00 & 0.090 &  0.0715 \\
HD 16175 b      &    990.00 &  0.60 &  222.00 & 0.082 &  0.0074 \\
HD 3651 b       &     62.22 &  0.60 &  245.50 & 0.083 &  0.2431 \\
HD 190984 b     &   4885.00 &  0.57 &  318.00 & 0.085 &  0.0014 \\
HIP 2247 b      &    655.60 &  0.54 &  112.20 & 0.159 &  0.0035 \\
HD 175167 b     &   1290.00 &  0.54 &  325.00 & 0.101 &  0.0051 \\
HD 190228 b     &   1136.10 &  0.53 &  101.20 & 0.142 &  0.0010 \\
HD 87883 b      &   2754.00 &  0.53 &  291.00 & 0.113 &  0.0033 \\
HD 142022 b     &   1928.00 &  0.53 &  170.00 & 0.102 &  0.0025 \\
HD 108147 b     &     10.90 &  0.53 &  308.00 & 0.100 &  1.3664 \\
HD 168443 b     &     58.11 &  0.53 &  172.95 & 0.097 &  0.0988 \\
HD 81040 b      &   1001.70 &  0.53 &   81.30 & 0.171 &  0.0016 \\
HIP 5158 b      &    345.72 &  0.52 &  252.00 & 0.119 &  0.0241 \\
HD 4203 b       &    431.88 &  0.52 &  329.10 & 0.104 &  0.0127 \\
HD 217107 c     &   4270.00 &  0.52 &  198.60 & 0.100 &  0.0012 \\
HAT-P-2 b       &      5.63 &  0.52 &  185.22 & 0.100 &  2.5595 \\
HD 1237 b       &    133.71 &  0.51 &  290.70 & 0.117 &  0.0650 \\
HD 142415 b     &    386.30 &  0.50 &  255.00 & 0.132 &  0.0170 \\
HD 34445 b      &   1000.00 &  0.49 &  131.00 & 0.145 &  0.0025 \\
HD 215497 c     &    567.94 &  0.49 &   45.00 & 0.150 &  0.0048 \\
HD 106252 b     &   1531.00 &  0.48 &  292.80 & 0.135 &  0.0045 \\
HD 33636 b      &   2127.70 &  0.48 &  339.50 & 0.117 &  0.0025 \\
HD 33283 b      &     18.18 &  0.48 &  155.80 & 0.120 &  0.2591 \\
HD 196885 b     &   1333.00 &  0.48 &   78.00 & 0.179 &  0.0010 \\
HD 181433 d     &   2172.00 &  0.48 &  330.00 & 0.119 &  0.0030 \\
HD 210277 b     &    442.19 &  0.48 &  119.10 & 0.190 &  0.0046 \\
HD 154857 b     &    409.00 &  0.47 &   59.00 & 0.187 &  0.0039 \\
HD 187085 b     &    986.00 &  0.47 &   94.00 & 0.238 &  0.0014 \\
HD 147018 b     &     44.24 &  0.47 &  335.97 & 0.118 &  0.1681 \\
HD 66428 b      &   1973.00 &  0.47 &  152.90 & 0.130 &  0.0015 \\
HD 50554 b      &   1224.00 &  0.44 &    7.40 & 0.137 &  0.0031 \\
HD 23127 b      &   1214.00 &  0.44 &  190.00 & 0.135 &  0.0035 \\
HD 202206 b     &    255.87 &  0.44 &  161.18 & 0.146 &  0.0114 \\
HD 74156 c      &   2473.00 &  0.43 &  258.60 & 0.156 &  0.0022 \\
4 UMa b         &    269.30 &  0.43 &   23.81 & 0.153 &  0.0096 \\
HD 213240 b     &    882.70 &  0.42 &  201.00 & 0.144 &  0.0048 \\
HD 117618 b     &     25.83 &  0.42 &  254.00 & 0.156 &  0.3417 \\
HD 141937 b     &    653.22 &  0.41 &  187.72 & 0.150 &  0.0058 \\
HD 65216 b      &    613.10 &  0.41 &  198.00 & 0.148 &  0.0071 \\
HD 126614 A b   &   1244.00 &  0.41 &  243.00 & 0.162 &  0.0042 \\
70 Vir b        &    116.69 &  0.40 &  358.71 & 0.149 &  0.0307 \\
HD 171238 b     &   1523.00 &  0.40 &   47.00 & 0.182 &  0.0015 \\
HD 5388 b       &    777.00 &  0.40 &  324.00 & 0.155 &  0.0055 \\
HD 11977 b      &    711.00 &  0.40 &  351.50 & 0.154 &  0.0041 \\
HD 181433 b     &      9.37 &  0.40 &  202.00 & 0.147 &  1.1056 \\
14 Her b        &   1754.00 &  0.39 &   19.60 & 0.163 &  0.0016 \\
HD 125612 b     &    510.00 &  0.38 &   21.00 & 0.180 &  0.0054 \\
42 Dra b        &    479.10 &  0.38 &  218.70 & 0.163 &  0.0091 \\
  \enddata
  \tablecomments{$\Delta t$ in units of orbital phase.}
\end{deluxetable}

The calculated values of $\epsilon_{\mathrm{max}}$ and $\Delta t$ for
$\sim 70$ of the most eccentric known exoplanets are tabulated in
Table \ref{phasetable}. Of the planets represented in this table, the
planet with the highest eccentricity, HD~80606b, is also the planet
with the highest predicted flux ratio. This is not surprising
considering that this planet's periastron passage is behind the star,
leading to the high secondary eclipse probability and the subsequent
observation of that eclipse by \citet{lau09}. As described earlier and
demonstrated by Figure \ref{knownexo}, the flux ratios of the planets
in Table \ref{phasetable} are dominated by the period and therefore
the semi-major axis of the orbits.


\section{Orbital Parameter Refinement}

As described by \citet{kan09b}, the refinement of orbital parameters
is not only an essential component for succussful detection of
features which only appear for a small fraction of the orbit, it is
also achievable with relatively few additional radial velocity
measurements. This is particularly true of long-period planets whose
orbits tend to have higher associated uncertainties and for which
opportunities to observe at a particular place in the orbit are far
less frequent.

For the goal of attempting to detect a planetary transit, it is the
time of predicted transit mid-point which needs to be constrained. For
optimal observations of phase variations, it is the time span during
which the maximum change in flux ratio occurs which needs to be
accurately determined, previously defined by the quantity $\Delta
t$. The reason for this is because, even though the phase variation
occurs over the entire orbit, it is assumed that the high cadence and
precision needed will necessitate limited observing time using
highly-subscribed instruments (discussed further in Section
\ref{casestudies}).

\begin{figure}
  \includegraphics[angle=270,width=8.2cm]{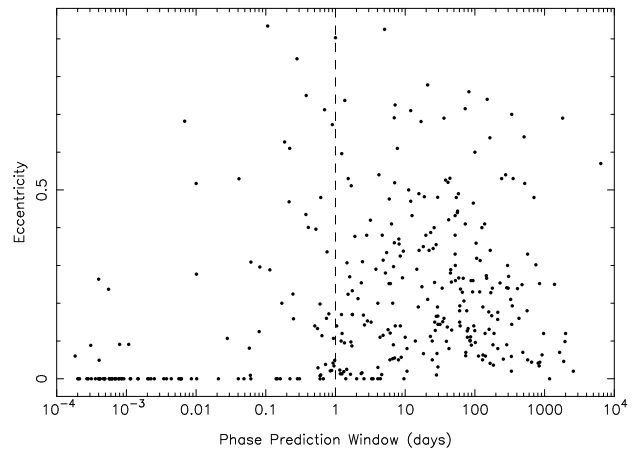}
  \caption{Orbital eccentricity as a function of phase prediction
    window for the known exoplanets. The dashed vertical line
    indicates a reasonable boundary beyond which observational
    attempts during a predicted phase location become difficult.}
  \label{phasewin}
\end{figure}

Using the analogy of the transit window described by \citet{kan09b},
we introduce the concept of the phase prediction window, which is the
time period during which a particular phase of the orbit could occur
according to the uncertainties associated with the orbital period and
the time of periastron passage. In Figure \ref{phasewin} we plot
orbital eccentricity as a function of the phase window for the 370
exoplanets for which the necessary uncertainties were available
(358). It is clear that the planets with the highest eccentricities
are the most difficult cases to predict orbital phase locations. This
is important because the most eccentric orbits tend to have much
smaller $\Delta t$ values and so it is essential that orbital
refinement be used to reduce the uncertainty in the phase prediction
and hence the size of the associated window.


\section{Robustness of Phase Models}
\label{robustness}

Although the phase function and albedo formulation adopted in this
study to compute the expected phase variations are physically
motivated, the sample of exoplanets with accurate measurements for
these functions is relatively small. For example, the Hilton phase
function (Equation \ref{phase}) is based upon the cloud maps of Venus
and Jupiter which, although they produce similar phase functions, have
their own unique cloud configurations with resulting slight variations
in their respective phase functions. Likewise, real exoplanets could
potentially exhibit a range of phase functions and albedo
distributions beyond those considered here. As is clear from Equation
\ref{fluxratio}, the flux ratio is linearly dependent on the geometric
albedo. In other words, a 1\% change in the value of $A_g$ translates
into a 1\% change in the value of $\epsilon_{\mathrm{max}}$. The
geometric albedo has no effect on the time between minimum and maximum
flux, $\Delta t$, since the shape of the phase variation is not
altered.

\begin{figure}
  \includegraphics[angle=270,width=8.2cm]{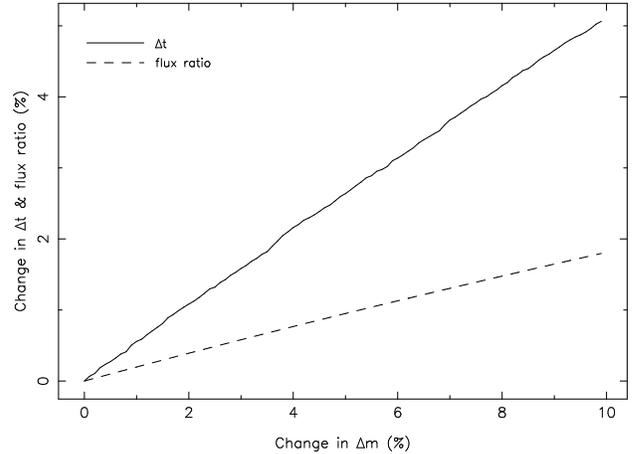}
  \caption{The percentage change in the values of $\Delta t$ and
    $\epsilon_{\mathrm{max}}$ as a function the percentage variation
    in the planetary visual magnitude correction (see Equation
    \ref{magcorr}) for the planets in Table \ref{phasetable}.}
  \label{robustfig}
\end{figure}

A change in the phase function however, will alter both the calculated
values of $\Delta t$ and $\epsilon_{\mathrm{max}}$. In order to
quantify this effect, we varied the value of the planetary visual
magnitude correction, $\Delta m$, as described in Equation
\ref{magcorr}. This was performed for all of the planets shown in
Table \ref{phasetable}, from which the the mean of the percentage
change in $\Delta t$ and $\epsilon_{\mathrm{max}}$ was calculated for
those planets. The results of this simulation are shown in Figure
\ref{robustfig}. Even a substantial change in the phase function of
10\% leads to a relatively minor change in the expected value of
$\Delta t$ and an even smaller impact on the expected flux
ratio. Additionally, the magnitude of these changes are largest for
eccenctric planets which are more sentitive to the location of the
peaks in the phase curve.


\section{Case Studies and Detectability}
\label{casestudies}

Here we present specific examples of predicted phase amplitudes and
detectability for several of the known exoplanets. In evaluating
whether these signatures are detectable or if instead calculating
these signatures will remain a theoretical exercise for the immediate
future, consider the precision of the Hubble Space Telescope (HST) and
the MOST satellite. Both of these telescopes have observed the $V =
7.65$ star HD~209458. The HST observations by \citet{bro01} achieved a
precision of $1.1 \times 10^{-4}$ and the MOST observations by
\citet{cro07} achieved a precision of $3.5 \times 10^{-3}$. In these
cases, the necessarily high cadence resulting from the brightness of
the host stars could be used to advantage by binning the data to
produce higher precision. In addition, the ellipsoidal variations
detected by \citet{wel10} using Kepler data is of amplitude $3.7
\times 10^{-5}$.


\subsection{The HD~37605 System}

Our current knowledge of the HD~37605 system consists of a single
planet in a $\sim 54$ day, highly eccentric orbit \citep{coc04}. This
places this planet close to the top of the list shown in Table
\ref{phasetable}. The peak flux ratio from this planet is expected to
be $0.56 \times 10^{-5}$, more than a factor of 10 smaller than that
for HAT-P-7b. Even so, the peak flux ratio is helped substantially by
the periastron argument of $\omega = 211\degr$ which places the
periastron passage close to the observer--star line-of-sight on the
far side of the star.

The minimum time between minimum and maximum flux ratio is 0.04 of the
orbital phase, or $\sim 2.0$ days. Conversely, the phase prediction
window for this planet at the time of discovery (2004) was $\sim 1.4$
days. At the time of writing, this window has since grown to $\sim 19$
days due to the $\sim 40$ orbits which have occurred since then.
Thus, the precision of the orbital parameters require further
improvement before a robust attempt to only observe the $\Delta t$
section of the orbital phase is made.


\subsection{The HAT-P-13 System}

The HAT-P-13 system \citep{bak09} is particularly interesting because
it consists of an inner transiting planet ($P = 2.92$ days) in a
circular orbit with an outer companion in a long-period ($P = 428.5$
days) eccentric orbit. Additional data acquired by \citet{win10}
provide evidence for a third body in the system. The coplanarity of
the system remains in question however since the refined orbital
parameters for the outer planet have not yet resulted in a transit
detection. Could this be resolved with observations during the
predicted maximum flux ratio occurance of the outer planet? This will
be difficult for the following reasons. Firstly, the maximum flux
ratio of the inner planet is predicted to be $3.11 \times 10^{-5}$,
compared to $0.017 \times 10^{-5}$ for the outer planet. Detecting the
signature of the outer planet would be a challenging detection task
even if the signature of the inner planet were not present. Secondly,
since the inclination of the outer planet is not known, the predicted
flux ratio becomes an upper limit, potentially making the detection
criteria even more dire. The substantiative dominance of the inner
planet over the phase signature makes this far too challenging a task
for any current instruments.


\subsection{The HD~82943 System}
\label{hd82943}

\begin{figure}
  \includegraphics[angle=270,width=8.2cm]{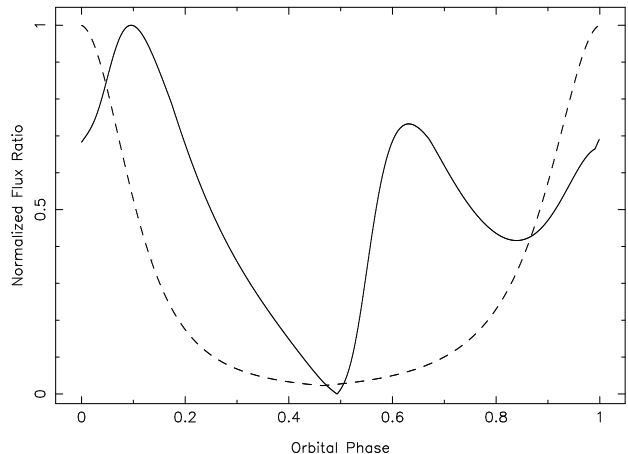}
  \caption{The predicted normalized flux ratio (solid line) for the
    HD~82943 system, the planets of which are in 2:1 resonance. The
    dashed line represents the phase function for the outer planet.}
  \label{hd82943fig}
\end{figure}

The HD~82943 system contains two planets in eccentric orbits in a
well-studied 2:1 resonance \citep{lee06}. The inner and outer planets
have periods of 219 and 441 days respectively but have near identical
maximum flux ratios of $\sim 0.008 \times 10^{-5}$; well beyond the
detection limits of current instruments. The reason for the similarity
in the peak flux ratios is due to the orbital elements in the system.
The eccentricity of the orbits combined with the resonance has led to
the axes of the orbits being almost $\pi$ out of phase with each
other. This leads to the complex structure for the total flux ratio
variation of the system, shown in Figure \ref{hd82943fig}. Note that
the periastron passage of the outer planet occurs behind the star,
whereas it occurs in front of the star for the inner planet, yielding
a relative increase in the maximum flux ratio for the outer planet,
partially compensating for the larger semi-major axis.


\section{Conclusions}

Current generation space-missions are already detecting exoplanet
phase variations in the optical (eg., Kepler) and the IR (eg.,
Spitzer). The steps these produce towards characterizing the
atmospheres of these exoplanets are significant since they provide
direct measurements of the atmospheric albedo and thermal
properties. This has currently been primarily undertaken for
short-period planets since many of these transit and produce phase
varations on easily observable timescales. However, current radial
velocity surveys are biased towards planets whose orbits are closer to
edge-on, since larger semi-amplitude signatures are produced, and
therefore biased towards planets with larger predicted phase
amplitudes.

We have shown here how time and position dependent functions for the
geometric albedo and phase can be used to describe expected phase
variations for long-period eccentric giant planets. There is a clear
degeneracy with orbital inclination and resonace when considering
multi-planet systems and care must be taken to account for these
possibilities. Applying these results to the known exoplanets shows
that many long-period eccentric planets can have significant peak flux
ratios, comparable to those of short-period planets. Additionally, the
phase prediction windows of eccentric planets, during which
observations will be optimally placed, will tend to be poorly
contrained. The refinement of orbital parameters for the known
exoplanets is clearly a key component for optimal observations of the
eccentric planets during maximum phase amplitude. Improving the
measured orbits of long-period planets is already being undertaken by
such projects as the Transit Ephemeris Refinement and Monitoring
Survey (TERMS) \citep{kan09b}. However, most of the predicted flux
ratios for the known planets push heavily against the boundaries of
what is achievable with current ground and space-based instruments. A
thorough search of all these planets will therefore likely need to
await future generation telescopes, such as the European Extremely
Large Telescope (E-ELT), the Thirty Meter Telescope (TMT), the Giant
Magellan Telescope (GMT), and the James Webb Space Telescope (JWST).

As more science results are released by the Kepler mission, the study
of photometric phase variations of long-period planets will become
increasingly relevant. Not only is it an existing mission which has
already detected phase variations in the light curve of HAT-P-7b, it
is specifically looking for transiting planets where there is an a
priori knowledge that the orbital inclination is favorable towards
maximum phase amplitude. Additionally, Kepler will eventually detect
transiting long-period ($P > 100$ days) planets where the bias will
certainly be towards eccentric orbits since those have a higher
probability of transiting \citep{kan08}. These future discoveries will
be prime candidates to detect the phase variations described here.


\section*{Acknowledgements}

The authors would like to thank David Ciardi for several useful
discussions. We would also like to thank the anonymous referee, whose
comments greatly improved the quality of the paper. This research has
made use of the Exoplanet Orbit Database and the Exoplanet Data
Explorer at exoplanets.org.


\end{document}